\documentclass[11pt,a4paper]{article}
\usepackage[hyperref]{emnlp-ijcnlp-2019}
\usepackage{times}
\usepackage{latexsym}

\usepackage{algpseudocode}
\usepackage{algorithm}
\usepackage{amsmath}
\usepackage{amssymb}

\usepackage{url}

\aclfinalcopy

\title{DimensionRank: Personal Neural Representations for Personalized General Search}

\author{Greg Coppola \\
Propaganda Gold \\
{\em greg@propaganda.gold }
}

\date{}

\begin{document}
\maketitle
\begin{abstract}
{\it Web Search} and {\em Social Media} have always been two of the most important applications on the internet.
We begin by giving a unified framework, called {\em general search}, of which which all search and social media products can be seen as instances.

{\bf DimensionRank} is our main contribution.
This is an algorithm for {\em personalized} general search, based on neural networks.
DimensionRank's bold innovation is to model and represent each user using their own unique personal neural representation vector, a learned representation in a real-valued multidimensional vector space.
This is the first internet service we are aware  of that to model each user with their own independent representation vector.
This is also the first service we are aware of to attempt personalization for general web search.
Also, neural representations allows us to present the first Reddit-style algorithm, that is immune to the problem of ``brigading''.
We believe personalized general search will yield a search product {\em orders of magnitude} better than Google's one-size-fits-all web search algorithm.

Finally, we announce {\em Deep Revelations}, a new search and social network internet application based on DimensionRank.

\end{abstract}

\section{Introduction}
{\it Web Search} and {\em Social Media} are two of the most important applications on the internet.
In the Western market, Google, Bing, Yahoo and Duckduckgo are notable web search products.
Facebook, Twitter, Instagram, Tumblr and Reddit are notable social media products.
YouTube is a notable product that contains characteristics of both search and social media.
Amazon is a mix of search and social media products, paired with real-world execution.
Airbnb is a kind of search and social media, paired with a supply of partnered tentants.
Spotify has elements of search and social media, paired with the licenses to a large catalog of songs.
Thus, we see that the areas of web search and social media are clearly important, because they impact the lives of billions of people, and because they produce very profitable companies.

We begin by providing a unified framework for {\em general search}.
This framework, firstly, helps us to analyze existing technology products.
Moreover, using this unified framework, we can create an {\em interdependent range} of products, each including elements of both search and social media, with each leveraging the same underlying user representations.

Our main contribution is {\bf DimensionRank}, a neural algorithm for personalized search.
DimensionRank's bold innovation is to model each user using their own unique personal neural representation, a learned vector in a real-valued vector space.
We further announce {\em Deep Revelations}, a new search and social network internet application based on DimensionRank.
This is the first service we are aware  of that to model each user with their own independent representation vector.
This is also the first service we are aware of to attempt personalization for general web search.
Finally, neural representations allows us to present the first Reddit-style algorithm, that is immune to the problem of ``brigading''.

\section{A Unified Framework for Web Search and Social Media}
We have observed that users typically think of ``{\em web search}" and ``{\em social media}" as two different categories of product.
This is probably because Google, as a complete product, is a very different from Facebook or Twitter, as complete products.
However, in another sense, both products based on the task of {\em information retrieval}:
\begin{quote}
     Information retrieval is finding material of an unstructured nature that satisfies an information need from within large collections. \citep{ir:manning:08}
\end{quote}

Since both web search and social media are just instances of information retrieval, it makes sense to create a unified framework that will encompass all of the modern notable products we have been discussing.
This way, we can see search and social media products as lying on a spectrum, or a product space.
Once we understand the contours of this product space, we can leverage this understanding to create new products inside the space.

\subsection{Keyword-Restricted Search vs. Keyword-Free Browsing}
Users will strongly associate Google with the text box where they type their search keywords.
In contrast, on Facebook or Twitter, one will typically thinking of just browsing their feed, without any search keywords.
However, Facebook and Twitter do in fact allow the user to add their own search terms, making it clear that a hard border between the two kinds of service is tenuous.
In fact, the materials posted on Facebook could probably be used to answer a good percentage Google web searches, but users don't think about Facebook as a search engine, probably because Facebook haven't optimized that side of their service, or emphasized it as such.
YouTube is an interesting example of a product where the user can either do keyword-free browsing, or do a keyword-restricted search.
In general, an information seeking session can begin with keywords, or not.
Whether keywords drive the session or not, we still have rich context about a user (e.g., preferences, time of day of the session, past session activity) to drive a search session.

\subsection{Machine-Crawled Ingestion vs. User Uploads}
In the early days of the internet, one of the central abilities that Google had was that a ablity to ``crawl'' the entire web.
Before the world's business was based on the internet, this ability to crawl the entire web was crucial, because Google search was not then important enough to motivate all web site owners to have bothered adding their sites to Google's index.
Also, in the early days of the internet, it was hard to find any sites on a topic at all.
The problem that Google was able to overcome was the scarcity of information.

Nowadays, many things have changed.
First of all, businesses and political entities are convinced of the importance of the internet.
Today, being absent from the internet is equated by many business leaders with having no existence at all.
And, the problem facing a user making a web search is that there will be {\em too many} matching documents for a query, rather than too few.

Deep Revelations is more interested in the latter task, of helping to make sense of topics in which there are {\em too many} results, and one has to rank them.
Requiring users to upload documents before they will be indexed acts as a kind of filter, and means that we are only storing documents that at least some users have already definitely shown that they like, by taking the time to upload posts about them.
Crawling is also the hardest part of traditional search to implement and maintain, so this is the first thing that we want to drop.

\subsection{Efficient Search vs. Leisure Browsing}
When using Google, users have developed the expectation that the correct answer is literally the first result, or else is on the first page.
Indeed, in 2005, then-CEO of Google Eric Schmidt called it a ``bug'' whenever the ``right answer'' was not exactly the first result, for any search.
Cf., ``We should be able to give you the right answer, just once. We should know what you meant... We should get it exactly right.. And we should never be wrong,'' \citep{rose:05}.
This is the epitome of the vision of ``efficient'' search.
This mode is characterized by the user's desire to stop the task of searching as soon as possible, preferably after the first result.

In contrast, when a user logs on to Facebook, Instagram or Reddit, they are usually expecting to {\em spend time} on the platform.
In other words, the goal is not that the user finish as fast as possible, but instead that they remain as long as possible.
YouTube is a notable example in which the user can either be looking for the first relevant search item, or else to browse.
This is a paradigm we want to pursue: give the user the option to either do an efficient search, or to do leisure browsing, with the same underlying database and prediction models serving both functions.
This way, we can leverage the signals that we get from users when they are browsing, to help predict what they will like when they are in a rush.

\subsection{Explicit vs. Implicit User Signals}
Users are quite familiar with the idea of ``liking" posts to show pleasure, on social networks like Facebook, Twitter or Instagram.
On YouTube and Reddit, you can even ``down vote" a post, indicating displeasure.
On the other hand, on Google search, there is no notion of voting either up or down.
Thus, a casual user might suppose that there is something inherent about search engines that mean that things should never be voted up or down.
In fact, this is another accident of history.
A search engine could request explicit signals.
We think that this could be a very powerful way to get signals for general search, and intend to use it.

\subsection{Personalized vs. Generic Results}
When a user logs on to Facebook, Twitter or Instagram, they will expect to see their own personal feed.
A generic, not personalized, social media feed arguably makes no sense.
However, when running a Google search, the user does not expect much personalization.\footnote{At least, the user {\em should not} expect this because, as we will see in \S \ref{s:google}, Google does not seem to implement much personalization.}
YouTube is a search engine that can produce personalized results as well.
Personalization is a main theme in this paper.
The fact that there is no personalized general search engine can be viewed, under this framework, as an accident of history, and one which we plan to change with DimensionRank.

\subsection{Relevance of Inter-User Connections}
In a canonical Facebook or Twitter session, a user sees only posts driven by their chosen connections.
The interpersonal links are the primary drivers behind recommendations.
In a Google search however, the relationship between the person who posted a link and the person searching is irrelevant.
The link must have been posted by the organization who owns the site.
This is rule is actually enforced by Google search.
So, some services are driven by inter-user connections, some are not, and some are probably somewhere in between.

\section{Algorithmic Analysis of Some Notable Services}
Different companies have released different amounts of information about their production algorithms.
For example, as we will review, Facebook and Bing have released open source code and written publicly available papers covering some aspects of their systems.\footnote{It is important to note that neither company releases enough details to completely replicate their production systems. The model descriptions are only partial.}
Other companies, like Google, do not give this much detail, but do release higher-level descriptions of their system to the public.
In this section, we will review what is publicly known about some notable services.

\subsection{Google}
\label{s:google}
Google is the canonical example of a keyword-restricted ``web search'' product.
As of 2015, Google says that their search engine contains a deep learning component, called {\em RankBrain} \citep{google:wired:15}.
RankBrain is in turn a ``signal" to a root-level search algorithm, along with ``thousands" of other signals.
Google also says that their root-level ranking model, which ultimately drives the ranking, is neither a deep learned model, nor a learned model at all.\footnote{
Cf., {\em ``The other signals, they're all based on discoveries and insights that people in information retrieval have had, but there's no learning," Greg Corrado, a senior research scientist at Google, told Bloomberg}  \citep{google:wired:15}.
}
Thus, it would seem that Google does not do personalization using personal user embeddings.
Indeed, separate reports indicate that Google does not heavily personalize search results at all \citep{schwartz:18}.

\subsection{Bing}
Microsoft has released as open source the code for their {\em Space Partition Tree And Graph} \citep{ChenW18}, which is one of the crucial algorithms behind Bing.
This model is very interesting.
But, it is more of a filtering algorithm than a ranking algorithm.
For ranking, it seems that Bing uses a machine learned model derived from their {\em learning to rank} framework \citep{burges:05,burges:10}.
Overall, it seems that Bing does in fact use deep learning for ranking, but does not attempt individual-level personalization.

\subsection{YouTube}
YouTube is considered the second biggest ``search engine'' in the world behind, Google search itself.
Since YouTube is personalized, but Google is not personalized, this makes YouTube the biggest personalized search engine in the world.
As discussed on \cite{fridman:youtube:20}, for personalized recommendations, YouTube uses a version of {\em collaborative filtering}.
Documents are put into clusters, according to whether they are likely to be watched by the same users.
Users are put into clusters according to whether they like the same documents.
These clusters are then used to suggest new recommendations, by identifying similar documents to what a user liked, or documents liked by similar users.
Thus, YouTube does provide a personalized service, but personalization happens using collaborative filtering, rather than creating a neural representation for the user.

{\em Collaborative Filtering Drawback} One drawback of this kind of collaborative filtering is that it users can be repeatedly led back to the same few {\em popular} documents.
For example, any time the present author starts listening to music on YouTube, the follow-on recommendations are always taken from the same set of well-known stars, over and over again.
The author would prefer to be recommended new music that he hadn't heard before.
This effect can happen in collaborative filtering if, for example, the system is set to be cautious about recommending new documents, and only recommends a document if multiple similar users like that document.
This will tend to promote popular songs, because more users in general like popular songs.
Collaborative filtering has a hard time distinguishing a gem that only a few people will know, versus an obscure song that people did not like.
Neural networks do not have to suffer from this same problem, because new songs are not identified using clusters, but instead by representing each document and each user with a unique neural representation.
These representations can be used to characterize the relationships between users and documents, and so to understand whether a document really will be of mutual interest.

\subsection{Facebook}
Facebook have described a system they call the {\em Deep Learning Recommendation Model (DLRM)} \citep{facebook:dlrm:17}.
This model uses deep learning to produce recommendations.
The the model is trained using {\em matrix factorization}.
It seems that the DLRM likely does not give each user its own unique representation vector, but we are not completely sure.
Facebook say that they let ``users and products be described by many continuous and categorical features'' (page 3).
This presumably means that a user is represented by a conjunction of a set of atomic features, shared accross users.
DLRM is probably the  most similar past work to DimensionRank.
There are, as we understand it, two major differences between DLRM and DimensionRank.
First, DimensionRank gives each user their own personal neural embedding representation, which DLRM doesn't seem to do, and certainly does not emphasize.
In contrast, for DimensionRank, the personal neural representation is the defining characteristic of our algorithm, and we have emphasized this repeatedly.
Second, DimensionRank emphasizes personalized {\em general search}, rather than simply emphasizing personalized recommendations for browsing, as Facebook do.
Also, we describe a concrete algorithm for personalized search, where Facebook do not.

\subsection{Airbnb}
Airbnb's personalized search task is to match a suitable visitor with a suitable host.
Both visitor and host must like each other.
Personalization is important on the site because not all users are mutally compatible.
\cite{airbnb:18} propose to use user embeddings for personalization, but not a unique representation for each user.
They represent each user as a ``type'' and a collection of a small number of atomic features, noting that ``storing embeddings for each user to perform online calculations would require lot of memory.''
Our work is quite similar to this, but differs in that we are giving each user their own unique representation vector, and that we are doing general search.

\subsection{Reddit and the Brigading Attack}
\subsubsection{Algorithm}
Reddit's algorithm is extremely simple, and quite powerful.
Each user can vote a post {\em up} or {\em down}  \citep{medium:reddit:15}.
Each post then has a score that is its number of up votes, minus its down votes, with some accounting for post freshness.
This simple system can be very effective in surfacing interesting posts to the ``front page".

\subsubsection{The Brigading Attack}
However, the Reddit algorithm has always been vulnerable to an attack called the {\em brigading} attack.
The attack must be carried out by a co-ordinated group of attackers, $\{a \in A\}$.
The target of the attack is a {\em community channel} (``subreddit'').
The channel will represent a certain community, $\{c \in C\}$.
In the Reddit algorithm, members of $A$ down-vote ``good posts'' from $C$.
Here, a {\em good post} for $C$ would be a post that most members $c \in C$ would up-vote, if they got to see it.
If enough members of $A$ downvote a post $d$ from $C$, the running score of $d$ can be pushed below a certain {\em threshold}, at which point the Reddit algorithm will remove $d$ from the running, before members of $C$ can see it.
\subsubsection{Possible Remedies}
The central problem with the Reddit algorithm is that it cannot determine who is in $A$ vs. $C$.
The Reddit algorithm has no notion of a predicate over users that it applies.
There is no mathematical difference in the way members of $A$ are treated from those of $C$.

{\em Hard Group Membership}
One solution to the brigading problem is to require users to get permission from a group owner in order to join.
This way, brigaders can simply be hardly excluded.
The drawback of this solution is that it loses the open nature of a Reddit channel.
A crucial property of Reddit is that any genuine member is free to join and start contributing to a community without having to know anyone first.
Open members supports the dynamic discover of inter-user connections on Reddit.

{\em Statistical Methods}
Another approach would be to use statistical methods.
We can imagine using statistics to identify clusters representing $A$ and $C$, and treat them differently.
This is effectively what we are going to do with embedding models in DimensionRank.
Embedding models {\em implicitly} cluster the data \cite{mikolov:13}.
This way, we can use the clusters to keep attackers separated from innocent communities, without having to impose hard the group membership constraints that destroy dynamic connection discovery.

\section{DimensionRank Overview}

{\bf DimensionRank} is an algorithm for general personalized search based on unique, independent neural representations for each user and document.
The central points of the strategy are as follows.

\subsection{Deep Learning}

We use deep learning.
Deep learning is a kind of machine learning, in which the learning process can discover its own representations of the input space.
The use of deep learning transfers the problem of algorithm creation.
Traditionally, creating a ranking algorithm meant hiring a small army of software engineers to directly write that ranking algorithm.
The use of machine learning and deep learning means that are task is simply to collect and label data.
Given labeled data, the learning algorithm can create its own rules and heuristics for ranking the results.
This means, using deep learning, we can create a high-quality ranking product with a far smaller team than would have been needed a generation ago.

\subsection{Personal Neural Representation for Each User}
The central bold innovation of our plan is to allow each user, and each document, their own personal neural embedding.
That is, each user is represented as a vector in $\mathbb{R}^n$, each document by a vector in $\mathbb{R}^m$, for some chosen values of $n$ and $m$.
Airbnb note that they would not give each user their own embedding because this would be too costly.
We will overcome the engineering challenges requried to give each user their own neural representation.	
This will be, we believe, the key differentiator for DimensionRank.

\subsection{Explicit User Signals}
We place a strong emphasis on {\em explicit}, rather than {\em implicit}, user signals.
That is, the user will tell us whether they liked or regretted a post, and how much.
Google famously hired an army of expensive and talented engineers to model user ``signals'' based on user behavior \citep{google:wired:15}.
This is expensive, time-consuming and, ultimately in our view, often less accurate than explicit signals.
We do not want to hire engineers to write algorithms guess the user's hidden itent.
Instead, one of our main design objectives is to create a full-stack system in which we can ingest as many {\em explicit} labels as possible.

\subsection{Stochastic Gradient Training}
For training, we will keep things as simple as possible, and will train the system with stochastic gradient descent.
This is a technical point, but we think it is important to emphasize.
Facebook discuss a matrix factorization approach \citep{facebook:dlrm:17}.
The choice of training method is not directly important to the user, and so we emphasize that this point is highly techincal, and not of primary importance to our users.
However, we believe, based on experience, that stochastic gradient descent will be easier for more developers to understand and work with.

\subsection{Interdependent Search and Social Media Products}
Finally, a central element of our strategy is to create multiple related information retrieval tasks, across which the user has the same representation.
For example, browsing-style sessions and products can be used to ingest web sites, and to get an idea about who a user is.
This knowledge can later be leveraged for efficient search, in which the user does not want to see as many documents.

\section{DimensionRank Algorithms}
We now look at the core algorithms behind DimensionRank.
We begin by discussing labeling and training, because our labeling and training flow is broadly the same, no matter what kind of information retrieval task the user is doing.
We then discuss two prediction cases: recommendations and search.

\subsection{Labeling}
The training pipeline begins when the user {\em labels} a post.
As we have said, we focus on the use of {\em explicit} user signals, meaning the user is going to explicitly tell us whether they liked to see a certain post, or not.
They do this by, for example, clicking on a button, or else by saying in the affirmative with their voice.
Also, some implicit signals will have value, and an implicit label is treated algorithmically the same way as an explicit label.
\begin{algorithm}[t]
\begin{algorithmic}[1]
\Function{ReceiveLabel}{$u, d, c, l$}
\Statex \Comment $u$ is the user
\Statex \Comment $d$ is a document
\Statex \Comment $c$ is the rest of the context
\Statex \Comment $l$ is a {\em label}
\State $e \leftarrow Example(u, d, c, l)$
\State add $e$ to $Q_{train}$
\EndFunction
\end{algorithmic}
\caption{\label{alg:label} Adding a Labeled Training Example}
\end{algorithm}
In Algorithm \ref{alg:label}, we depict the function \textsc{ReceiveLabel}, which receives a new user label and records a corresponding training example.
The example stores the subjective user $u$, the document $d$, other context $c$, and the label $l$ that we have just received from the user.
These are all stored in the example $e$ and serialized to disk.
Then, we add the identifier for $e$ to the training queue $Q_{train}$.
This example will be picked up asynchronously by the training server.

\subsection{Training}
Algorithm \ref{alg:train} depicts the training server.
This server simply loops forever, each time waiting for a single example from $Q_{train}$.
Training is purely {\em stochastic gradient descent (SGD)}.
We like SGD for its simplicity, proven record for accuracy, and its natural fit for handling {\em streams} of data.
On each round of training, we update the shared network weights $W$, as well as the object-specific embeddings weights for the user $u_e$, document $d_e$, and context $c_e$ variables, active in example $e$.
\begin{algorithm}[t]
\begin{algorithmic}[1]
\Function{TrainingServer}{$ $}
\Statex \Comment $Q_{train}$ is a queue of examples
\Statex \Comment $W$ are the network weights
\Statex \Comment $u_e, d_e, c_e$ are embedding parameters
\Loop
	\State wait for example $e$ from $Q_{train}$
	\State do forward pass for $e$
	\State do backward pass for $e$
	\State update embeddings for $u_e$, $d_e$, $c_e$
	\State update $W$
\EndLoop
\EndFunction
\end{algorithmic}
\caption{\label{alg:train} Training Server}
\end{algorithm}

\subsection{Recommendations for Leisure Browsing}
\begin{algorithm}[t]
\begin{algorithmic}[1]
\Function{RecommendationServer}{$ $}
\Statex \Comment $M$ is a trained model
\Statex \Comment $Q_{new}$ is a queue of newly posted documents
\Statex \Comment $Q_u$ is $u$'s personal recommendations
\Loop
	\State wait for document $d$ from $Q_{new}$
	\For{each user $u$}
		\If{$M$ predicts $u$ will like $d$}
			\State add $d$ to $Q_u$
		\EndIf
	\EndFor
\EndLoop
\EndFunction
\end{algorithmic}
\caption{\label{alg:feed} Personalized Recommendations}
\end{algorithm}

Assume that users are making new {\em posts}, each potentially linking to some {\em documents} on the web, to the system.
Each time a new post is made, it is stored to the database, and its identifier is added to $Q_{new}$.
That is, $Q_{new}$ is a queue containing each post is put onto when it is new, in order to be sorted by the system.
Algorithm \ref{alg:feed} shows the behavior of the recommendation server.
This algorithm assumes a personalized model $M$ for predicting whether a user $u$ will like a document $d$.
On each round, the server waits for a document from $Q_{new}$.
For each incoming document $d$, we will check whether each user $u$ wants to see $d$.\footnote{If this full calculation is not possible, we will approximate it.}
If the model believes that $u$ will want to see $d$, then $d$ is added to $Q_u$, the queue of recommendations for user $u$.
$u$ will receive recommendations from $Q_u$ when they next log on.
Thus, each user $u$ is receiving personalized recommendations, based on what $u$ in particular is expected to like.

{\em Dynamic Discovery of New Connections} Note that users do not need to explicitly form inter-user connections in order for this algorithm to work.
As written, any user can be shown a post by any other user, allowing for dynamic interpersonal connection discovery.
In practice, we can probably not ever afford to loop over all users looking for consumers for each document.
Thus, we will need to prune the search space, effectively perhaps reimposing a notion of ``connections''.
However, pruning should be done in a way to continue to promote this central value of Deep Revelations, which is the ability for the algorithm to dynamically discover like-minded users and connect them.

\subsection{Ranking for Keyword-Restricted Search}
Given the ability to make personalized recommendations, we can leverage this ability to also create personalized general search.
The mechanism behind this is very simple.
We employ a two-pass approach, as depicted in Algorithm \ref{alg:search}.
In the first pass, we rely on \textsc{GenericSearch}, a sub-routine that can retrieve high-scoring, but not personalized results.
For example, MongoDB has a search feature built into it.
Or, one can use a more advanced generic search solution like \citep{ChenW18}.
The results from this generic search are then re-ranked using a personalized model, to produce a personalized search result.

\begin{algorithm}[t]
\begin{algorithmic}[1]
\Function{KeywordSearch}{$k, u, c$}
\Statex \Comment $k$ are the keywords
\Statex \Comment $u$ is the user
\Statex \Comment $c$ is the rest of the context
\State $r \leftarrow GenericSearch(k)$
\Statex \Comment $r$ is {\em not} personalized
\State $R \leftarrow Personalize(r, u, c)$
\Statex \Comment $R$ is personalized for $u$ in $c$
\State {\bf return} $R$
\EndFunction
\end{algorithmic}
\caption{\label{alg:search} Personalized Keyword Search}
\end{algorithm}

\subsection{Hybrid Algorithms}
We can create hybrid derivatives using these basic building blocks.
Again, our intent is to create interdependent browsing and efficient search products, and leverage shared user representations between the two.

\section{Real-World Implementation}
Our own first implementation of DimensionRank is embedded into a new social network called {\em Deep Revelations}.
Deep Revelations is envisioned to be a next-generation Reddit, incorporating advances from deep learning and personalization.
The project can be found online at the URL:
\[
\href{https://deeprevelations.com}{deeprevelations.com}
\]
Links to the latest version of the source code will be found there.
We expect to launch the invite-only {\em alpha} version of the service in June 2020.

\section{Conclusion}
We have presented a unified framework, {\em general search}, for analyzing both traditional search and traditional social media products.
Our main contribution is {\bf DimensionRank}, a neural algorithm for personalized search.
DimensionRank's bold innovation is to model each user using their own unique personal neural representation, a learned vector in a real-valued vector space.
We believe this is the first proposal to either i) give each user their own personal neural representation vector, or ii) to propose highly personalized general search using deep learning.

\bibliography{emnlp-ijcnlp-2019}
\bibliographystyle{acl_natbib}

\end{document}